\documentclass[11pt]{article}

\usepackage{naacl2021}

\hypersetup{pdfproducer={XeLaTeX},%
            pdfauthor={Ilnar Salimzianov},%
            pdftitle={A baseline model for computationally inexpensive speech recognition for Kazakh using the Coqui STT framework},%
            pdfcreator={XeLaTeX}
}

\ifdefined\directlua
  \usepackage{fontspec}
\else
  \usepackage[T1]{fontenc}
  \usepackage[nomath]{lmodern}
\fi
\usepackage{times}
\usepackage{booktabs}
\usepackage{latexsym}
\usepackage{natbib}
\usepackage{hyperref}
\usepackage[hyphenbreaks]{breakurl}

\usepackage{comment}
\usepackage{subcaption}
\usepackage{multirow}

\usepackage{microtype}

\usepackage{graphicx}

\usepackage{ulem}

\title{A baseline model for computationally inexpensive speech recognition for Kazakh using the Coqui STT framework}

\author{Ilnar Salimzianov\\
  Taruen \\
  \texttt{ilnar@selimcan.org} \\
}

\begin{document}

\maketitle

\begin{abstract}
Mobile devices are transforming the way people interact with computers, and
speech interfaces to applications are ever more important. Automatic Speech
Recognition systems recently published are very accurate, but often require
powerful machinery (specialised Graphical Processing Units) for inference,
which makes them impractical to run on commodity devices, especially in
streaming mode. Impressed by the accuracy of, but dissatisfied with the
inference times of the baseline Kazakh ASR model of
\cite{khassanov-etal-2021-crowdsourced} when not using a GPU, we trained a new
baseline acoustic model (on the same dataset as the aforementioned paper) and
three language models for use with the Coqui STT framework. Results look
promising, but further epochs of training and parameter sweeping or,
alternatively, limiting the vocabulary that the ASR system must support, is
needed to reach a production-level accuracy.
\end{abstract}

\section{Introduction}

\subsection{Rationale}

Smartphones are widespread, and speech interfaces to applications are becoming
more and more important.

The performance of speech-to-text applications, as measured by word error rate
(WER) and character error rate (CER), is getting closer and closer to
0\%\footnote{\url{https://nlpprogress.com/english/automatic_speech_recognition.html}}.
However, best performing systems require powerful machinery (read: Graphical
Processing Units, GPUs) not found on commodity computers both for training
models (which is justifiable), but often also for inference, which makes them
impractical to run on low-power devices such as smartphones.

Often speech data is processed through APIs of big companies. At the same time,
companies having access to and collecting large amounts of sensitive data is
not without concerns, and many people, all other things being equal, would
prefer their speech-to-text or text-to-speech applications be libre/open-source
software that run locally, on \textbf{their} devices, without sending private
data off to someone else's server. Depending on the volume of speech/text data
to be processed, cost of using APIs can also be an issue.

Needless to say, state-of-the art automatic speech recognition (ASR) systems
are data-driven, and their accuracy is a function of the amount of speech data
available to train them. Fortunately, new datasets are constantly emerging. So,
in September 2020, a large speech corpus of Kazakh\footnote{A Turkic language
  mainly spoken in Kazakhstan and other Central Asian republics, China and
  Russia by about 13 million people \cite{ethnologue2021}}, available under a
Creative Commons Attribution 4.0 International
license\footnote{\url{https://creativecommons.org/licenses/by/4.0/}}, was first
presented \cite{khassanov-etal-2021-crowdsourced}\footnote{The corpus is
  available at \url{https://doi.org/10.48342/gkg9-gn84}. We deducted its first
  publication date from the following preprint:
  \url{https://arxiv.org/abs/2009.10334}}. As of June 2021, in version 1.1, the
corpus contains 332 hours of read speech and is, to our knowledge, the largest
speech corpus of Kazakh published. In addition, authors of the corpus trained
ASR models on it, and made them publicly
available\footnote{\url{https://github.com/IS2AI/ISSAI_SAIDA_Kazakh_ASR}}. We
wrote a simple web interface to the best performing model of
\cite{khassanov-etal-2021-crowdsourced} and packaged it into a Docker
image\footnote{Available at \url{http://taruen.com/hub.html}}. The model is
very accurate (indeed, if not in terms of word error rate (WER), at least in
terms of character error rate (CER) we consider it state of the art or very
close to it) but we weren't satisfied with inference times when not using a
GPU. It is not surprising that a deep learning-based model is relatively slow
when not utilising a special GPU, yet we wanted to be able to deploy the Kazakh
ASR system on commodity machines, including smartphones, and use it in
streaming mode.

\subsection{Objectives}

Thus the main objective of our study was to train and deploy an ASR system
known to be fast enough on computers without a specialised GPU, and see how it
performs in terms of accuracy and speed.

\section{Experimental}

\subsection{Acoustic model}

We trained a baseline acoustic model for Kazakh using the Coqui STT
    framework\footnote{\url{https://github.com/coqui-ai/STT}, version 0.9.3} on
    the corpus published by \cite{khassanov-etal-2021-crowdsourced} (version
    1.1, 332 hours of speech).

Except for the batch size, hyperparameters and number of training epochs were
identical to that described in the ``Baseline models'' section of
\cite{tyers-meyer2021shall}, which in turn were based on
\cite{ardila-etal-2020}. Concretely, version 0.9.3 of the English
model\footnote{\url{https://github.com/coqui-ai/STT/releases/tag/v0.9.3}}
served as the source model for transfer learning. We dropped 2 final layers of
it. Dropout was set to 0.05, learning rate to 0.001 and SpecAugment
\cite{Park2019} option was turned off. With these settings, we trained a model
for 25 epochs, without early stopping. Training was done on a single
\texttt{g1.1} machine of the Yandex Data Sphere
service\footnote{\url{https://cloud.yandex.com/en/services/datasphere}}. The
train, dev and test batch sizes was empirically set to 9, which kept the GPU
utilisation oscillating between approximately 75-95\%.

We made use of the scripts and the Docker file from the
\texttt{commonvoice-docker}
repository\footnote{\url{https://github.com/ftyers/commonvoice-docker/}} of
\cite{tyers-meyer2021shall} almost as is. We had to tweak the docker file
slightly to accommodate for details how the Yandex Datasphere works.

The train/dev/test split was kept as released by
\cite{khassanov-etal-2021-crowdsourced} (see Table 1 of the cited paper).

The \texttt{csv} files of the corpus were converted to the format Coqui STT
expects.
    
Besides this technical conversion, there was one more minor change:
approximately 20 cases were found where a Latin character (from Extended Latin
character set) was typed instead of a letter of the Cyrillic Kazakh alphabet
(as the rest of the transcriptions in the corpus are written
in)\footnote{e.g. U+0259 instead of the correct U+04D9}. All occurrences of
these Latin characters were replaced with corresponding Cyrillic
characters\footnote{There was also a company name which in other circumstances
  would be fine to spell with Latin characters, but for the reason explained
  above, we have re-written it in all-Cyrillic as well.}. The reason for this
change is that when training a Coqui STT model one has to specify the alphabet
(character set) that the resulting model should recognise, and transcriptions
should contain only the characters specified in the alphabet. Our target
alphabet consisted only of the letters of the Cyrillic Kazakh alphabet, so that
we had to pre-process the transcriptions of the corpus making sure that it
contains only Cyrillic Kazakh letters.

\subsection{Language models}

In ASR, an acoustic model can be and usually is complemented with a language
model (LM). An LM supported by Coqui STT can be trained effectively without a
GPU and independently from training the acoustic model using the \texttt{kenlm}
tool
\cite{heafield-2011-kenlm}\footnote{\url{https://kheafield.com/code/kenlm}}.
All in all we built and tested three language models. The first one was made
only on transcriptions from the \texttt{train} and \texttt{dev} sets of the
speech corpus (about 1.6 million tokens in total). In addition, two larger LMs
were constructed, which did \textbf{not} include any transcriptions from the
speech corpus so that no bias is created. The first of these larger LMs was
trained on the fiction texts scraped from the \url{kitap.kz} website on October
2017. The second was trained on the union of the same fiction texts with a
collection of news texts\footnote{Sources include \url{egemen.kz},
  \url{today.kz}, \url{akorda.kz}, \url{nur.kz} and several others.} and a
snapshot of Kazakh Wikipedia. Both corpora were pre-processed with the
\texttt{covo}
utility\footnote{\url{https://github.com/ftyers/commonvoice-utils}. Concretely,
  texts were piped through two commands: \texttt{covo segment kk} followed by
  \texttt{covo norm kk}} of \cite{tyers-meyer2021shall}. After pre-processing,
the fiction corpus contained about 20 million, the fiction+news+wikipedia
corpus about 44 million tokens.

\section{Results}

We evaluated the acoustic model, as well the acoustic model complemented with
either of the three language models just discussed\footnote{When ``packaging''
  a language model for using it with Coqui STT (let's call the resulting
  package a ``scorer'' to differentiate the two), it's possible to specify the
  so-called \texttt{default\_alpha} and \texttt{default\_beta} values. We had 2
  sets of those. The first were the \texttt{default\_alpha} and the
  \texttt{default\_beta} values hard-coded for all experiments of
  \cite{tyers-meyer2021shall}, hard-coded because it's assumed that these two
  values (if they are in a reasonable range) don't affect the accuracy by much
  to justify optimising them for all of the 31 languages that
  \cite{tyers-meyer2021shall} trained ASR models for. But to be on the safe
  side, we also let the \texttt{lm\_optimizer.py} script calculate the values
  of \texttt{default\_alpha} and \texttt{default\_beta} it deems
  optimal. Optimal \texttt{default\_alpha} and \texttt{default\_beta} were set
  to 1.2143912484271524 and 2.1012243193402487, respectively. The
  \texttt{default\_alpha}, \texttt{default\_beta} values of scorers of
  \cite{tyers-meyer2021shall} were 0.931289039105002 and 1.1834137581510284,
  respectively. Three language models times two sets of
  (\texttt{default\_alpha}, \texttt{default\_beta}) values resulted in 6
  scorers to evaluate. We had evaluated all 7 seven cases (i.e. acoustic model
  only and acoustic model combined with either of the 6 scorers), but there was
  no difference in results in terms of WER/CER when using the ``optimal''
  values versus values taken from \cite{tyers-meyer2021shall}, so we don't
  discuss them any further.} on the test set of
\cite{khassanov-etal-2021-crowdsourced} in terms of accuracy and processing
time.

All evaluations were run on a Lenovo Thinkpad T440p laptop\footnote{An Intel(R)
  Core(TM) i7-4700MQ CPU @ 2.40GHz with 12 GB of memory.} with the GPU
\textbf{disabled}. Test batch size was set to 8.
      
Table \ref{table:results} shows WER and CER for each of the models / model
combinations. In Table \ref{table:time}, time is shown of how long it took for
each evaluation to complete. The CTC loss of the acoustic model was 42.921391.

\begin{table}
  \centering
  \begin{tabular}{lrr}
    \toprule
    \textbf{Model} & \textbf{WER} & \textbf{CER} \\
    \midrule
    Acoustic only                     & 59.08 & 15.53 \\
    Acoustic + train-dev-set-lm       & 25.94 & 11.22 \\
    Acoustic + fiction-lm             & 34.30 & 14.58 \\
    Acoustic + fiction-news-wiki-lm   & 28.22 & 12.29 \\    
    \bottomrule
  \end{tabular}
  \caption{\textbf{Baseline results}. Word Error Rate (WER) and Character Error
    Rate (CER) of each of the models / model combinations on the held-out test
    set. The CTC loss of the acoustic model was
    42.921391.}\label{table:results}
\end{table}

\begin{table}
  \centering
  \begin{tabular}{lr}
    \toprule
    \textbf{Model} & Time \\
    \midrule
    Acoustic only                     & 2:23 \\
    Acoustic + train-dev-set-lm       & 0:46 \\
    Acoustic + fiction-lm             & 0:47 \\
    Acoustic + fiction-news-wiki-lm   & 0:43 \\    
    \bottomrule
  \end{tabular}
  \caption{\textbf{Processing time}. For each of the models / model
    combinations X, time (hours:minutes) is shown of how long it took for an
    evaluation of X on the held-out test set to complete. The main bulk of work
    during evaluation is transcribing each audio in the test set, and in the
    discussion below we use the above durations for estimating how long it
    takes to transcribe 1 second of speech using each of the models /
    combinations.}\label{table:time}
\end{table}

\section{Discussion}

The error rates observed are much higher than that of the best model of
\cite{khassanov-etal-2021-crowdsourced} (8.7\% WER and 2.8\% CER)\footnote{Keep
  in mind that the main goal of the authors of the cited article is not
  building an optimal ASR system but rather presenting a speech corpus, and
  models presented are initial exploratory models (and that authors use a
  different framework)} but in the range of what can be expected of 25 epochs
of training without tuning the parameters.

Recall that our objective was to train and deploy an ASR system known to be
fast enough on computers without a specialised GPU, and see how it performs in
terms of accuracy and speed, thus a few words must be said about the inference
time. The total duration of audio files in the test set was 7.1 hours or, to be
more precise, 25436 seconds. Since from Table \ref{table:time} we know how long
it took for each evaluation to run, and that each evaluation ran on 8 CPU
cores, we can estimate how long it would take to transcribe 1 second of speech
on a single CPU core by the following formula: [evaluation time in seconds] * 8
/ 25436. The estimates are as follows: 2.27 seconds per second of audio when
using the acoustic model only, and 0.87, 0.89 and 0.81 seconds per second of
audio when using the acoustic model combined with each of the three language
models\footnote{These numbers do not include the time to load the acoustic
  model and the scorer, but for the models in question on our laptop it does
  not take more than 0.01 seconds to load either.}. In short, we can conclude
that the model can be deployed for use in the streaming mode on commodity
machines and possibly smartphones.

The logs of the training script showed that the acoustic model had most likely
not converged after 25 epochs of training (which is to be expected), so we hope
that further training combined with parameter sweeping and the SpecAugment
feature will decrease the error rates. The reason for limiting ourselves to 25
epochs was mainly due to the cost of training. At the time we did not have
access to a GPU server of our own, but now we do, and are working towards
training for more epochs and finding optimal parameters.

\section{Related work and frameworks}

For an overview of works on Kazakh speech recognition and synthesis and
available speech corpora, we refer the reader to Section 2 of
\cite{khassanov-etal-2021-crowdsourced}.

As for more recent developments, it is worth mentioning that since January 2021
Common Voice \cite{ardila-etal-2020} -- a relatively new, multilingual dataset
with the goal of collecting speech data for all languages and releasing the
data into public domain -- also includes Kazakh. Common Voice releases have
been happening twice a year, and Kazakh is expected to land in the mid-year
relase of 2021, albeit the amount of data in Kazakh will probably be only
moderate by then.

On a related note, any non-English language is often called `low-resourced' in
the literature \cite{mika-2021}, but calling Kazakh low-resourced now that
there is a 332 hours-big freely available corpus (at least in the context of
speech recognition) would be an injustice to its authors and many other
researchers who have worked on Kazakh, so we refrain from calling it
that. Besides, transfer learning, pre-training and other methods of that kind
are blurring the distinction between what is ``low-resourced`` and what is
``high-resourced'' even more.

There are many ASR tool kits to choose from. If we consider our desiderata of
running ASR models on low-power devices, both
Vosk\footnote{\url{https://alphacephei.com/vosk/}} and Coqui STT would have
probably been equally valid choices. A more thorough comparative study of
frameworks is left for future work. Our experience with Coqui STT was that is
has a supportive community around it and an easy-to-follow documentation.

\section{Conclusion}

To our knowledge, we have presented the first ASR system for Kazakh usable on
commodity (read: non GPU) computers in streaming mode. Our acoustic model and
language models (scorers) can be downloaded from the following URL:
\url{https://drive.google.com/drive/folders/1OmME4_sy2-xW739fm7zRLr0cqyYS0ArY?usp=sharing}

\section{Acknowledgements}

This study was supported by the Nazarbayev University Research Program OPCRP2021014.

\bibliography{anthology,custom}
\bibliographystyle{acl_natbib}

\end{document}